\documentclass[prl,aps,twocolumn]{revtex4}

\usepackage{amssymb}
\usepackage{amsmath}
\usepackage{epsfig}

\begin{document}
\title{Entanglement of formation from optimal decomposition}
\author{Lin Chen}
\author{Yi-Xin Chen}
\affiliation{Zhejiang Insitute of Modern Physics, Zhejiang
University, Hangzhou 310027, China}

\begin{abstract}
We present a new method of analytically deriving the entanglement
of formation of the bipartite mixed state. The method realizes the
optimal decomposition families of states. Our method can lead to
many new results concerning entanglement of formation, its
additivity and entanglement cost. We illustrate it by
investigating the two-qubit state, the separable state, the
maximally correlated state, the isotropic state and the Werner
state.
\end{abstract}
\maketitle

The rapid progress in quantum information theory (QIT) has made
the entanglement the key resource in quantum mechanics. One of the
most fundamental questions in QIT is how to quantitatively
describe the entanglement by entanglement measures
\cite{Horodecki3}.

The main proposed entanglement measures so far include the
entanglement of formation (EOF) $E_f$ (and the corresponding
asymptotic generalization, the entanglement cost $E_c$) and
distillable entanglement $E_d$ \cite{Bennett1}. They respectively
present the entanglement of a state in terms of EPR pairs that is
required to create it and that can be extracted from it. All the
measures are equivalent to the von Neumann entropy
$E(|\psi\rangle)=S(\mbox{Tr}_A|\psi\rangle\langle\psi|)$,
$S(\rho)=-\mbox{Tr} \rho \ \mbox{log}\rho$ for pure bipartite
states \cite{Bennett2}.

The analytical calculation of entanglement measures is important.
The entanglement measure quantifies the quantum correlation in the
state used for any quantum-information task, e.g., an EPR pair is
both sufficient and necessary for quantum teleporation. A straight
way to prove an important conjecture that the formation of mixed
entanglement is irreversible \cite{Yang,Vollbrecht1}, is to
calculate the entanglement cost and distillable entanglement. If
the conjecture is true, it will mean that the mixed entanglement
manipulation is irreversible under local operations and classical
communications (LOCC) \cite{Vidal1}. On the experimental side,
there has been recent methods of estimating entanglement measures
like the EOF \cite{Walborn}. Entanglement measures have been also
shown to be useful in many other fields such as quantum phase
transition, spin-boson model and optical lattices \cite{Chen1}.

Despite the importance of entanglement measures, it is very
difficult to derive their analytical expressions for mixed
bipartite states by reason of the involved optimization problem.
Main progress was made in the derivation of EOF
\cite{Wootters,Terhal2,Vollbrecht2}. There are also a few results
on entanglement cost \cite{Horodecki2,Vidal2,Matsumoto} and
distillable entanglement \cite{Rains,Horodecki2}.

In this Letter we derive the EOF by means of the optimal
decomposition (OD) of states. We show that a known OD from a state
always leads to an OD family containing infinite number of states,
and the EOF of any state in it is computable. Next, new OD
families can be derived from the tensor product of OD families
with additive EOF, so we can continue to generate states with
computable EOF. Differing with the former skills, our method
mostly computes the EOF of states which are not symmetric and
whose subsystems have different dimensions. It greatly enlarges
the family of states whose EOF is computable and additive.
Moreover, Our method can evaluate the entanglement cost for many
states, while the latter is important but very difficult to
compute so far. The new results on additivity of EOF help get
insight into the classical capacity of quantum channels
\cite{Shor}.

We give the method of finding out the OD of a state with known
EOF. We determine the ODs of rank-2 two-qubit states with additive
EOF and two important states in QIT, the Werner state
\cite{Werner} and the isotropic state \cite{Horodecki4}. We
describe how the ODs of these states lead to computable EOF and
entanglement cost when the EOF is additive. We propose the schemes
for experimentally generating the Werner state and checking the
EOF. The separable states and maximally correlated (MC) states of
arbitrary dimension, whose EOF turns out to be additive, are also
analyzed. They provide strong evidence that the entanglement cost
is strictly larger than the distillable entanglement.

Let us start by recalling the definition of EOF. A mixed state
$\rho=\sum_ip_i|\psi_i\rangle\langle\psi_i|$ can be prepared in
many ways of ensembles of pure states $\{p_i,|\psi_i\rangle\}$.
The EOF for $\rho$ is defined in the way such that
\begin{equation}
E_f(\rho)=\underset{\{p_i,|\psi_i\rangle\}}{\mbox{min}}
\sum\limits_ip_iE(|\psi_i\rangle).
\end{equation}
Suppose the ensemble $\{q_i,|\phi_i\rangle,\forall q_i>0\}$
realize the EOF and it thus constitutes the OD of $\rho$. By
randomly changing the probabilities, we refer to the set of states
$\varepsilon_\rho$ constituted by the ensemble
$\{r_i,|\phi_i\rangle,\forall r_i\geq0\}$ as the OD family of
$\rho$. That is, a state $\sigma\in\varepsilon_\rho$ if and only
if $\sigma=\sum_ir_i|\phi_i\rangle\langle\phi_i|,r_i\geq0$. For
simplicity, we denote $\varepsilon_\rho=\{r_i,|\phi_i\rangle\}$ or
$\varepsilon_\rho=\{r_0,r_1,...,|\phi_0\rangle,|\phi_1\rangle,...\}$
throughout the paper. The EOF of any state in an OD family is
computable \cite{Vollbrecht2}, namely
$E_f(\sigma)=\sum_ir_iE(|\phi_i\rangle)$. In addition, the subset
of an OD family is also an OD family.

On the other hand, the OD family is closely connected to the
additivity of EOF of $\rho$. The latter means for any bipartite
state $\sigma$, it holds that $
E_f(\rho\otimes\sigma)=E_f(\rho)+E_f(\sigma).$ The problem that
whether EOF is additive is important \cite{Hayden}, for it will
lead to the equation $E_f=E_c$. The following facts easily follow
from the definition of additivity.

\textit{Lemma 1.} (1) The EOF of any state in the OD family of
$\rho$ is additive if the EOF of $\rho$ is additive; (2) If the
EOF of two states are additive, then the EOF of their tensor
product is additive.

\textit{Proof.} (1) Let the OD family of $\rho$ be
$\varepsilon_\rho=\{r_i,|\phi_i\rangle\}$. We are always able to
write $\rho=\sum_ip_i \rho_i$, with each state
$\rho_i\in\varepsilon_\rho$ and $\forall p_i>0$. For any state
$\sigma$ we get
\begin{eqnarray}
E_f(\rho)+E_f(\sigma)&=&\sum_ip_i(E_f(\rho_i)+E_f(\sigma))\nonumber\\
&\geq&\sum_ip_iE_f(\rho_i\otimes\sigma)\geq
E_f(\rho\otimes\sigma),
\end{eqnarray}
where we have used the fact that the EOF is subadditive and convex
\cite{Horodecki3} in the inequalities. Because $E_f(\rho)$ is
additive and the state $\rho_i$ is arbitrary, we conclude that the
EOF of any state in
$\varepsilon_\rho$ is also additive.\\
(2) Suppose $E_f(\rho_1)$ and $E_f(\rho_2)$ are additive. For any
state $\sigma$, it holds that $
E_f(\rho_1\otimes\rho_2\otimes\sigma)=E_f(\rho_1)+E_f(\rho_2\otimes\sigma)
=E_f(\rho_1\otimes\rho_2)+E_f(\sigma). $ So the EOF of the tensor
state $\rho_1\otimes\rho_2$ is additive. This completes the proof.

\textit{Proposition 1.} Let
$\varepsilon_{\rho_1}=\{p_i,|\psi_i\rangle\}$ and
$\varepsilon_{\rho_2}=\{q_i,|\phi_i\rangle\}$ be two OD families
and at least one of $E_f(\rho_1)$ and $E_f(\rho_2)$ be additive.
Then the set
$\varepsilon_{\sigma}=\{r_{ij},|\psi_i\phi_j\rangle\}$ is an OD
family. If both of $E_f(\rho_1)$ and $E_f(\rho_2)$ are additive,
then the EOF of any state $\sigma^{\prime}\in\varepsilon_{\sigma}$
is additive.

\textit{Proof.} Suppose the ODs of the states $\rho_1$ and
$\rho_2$ are constituted by the ensembles
$\{p^{\prime}_i,|\psi_i\rangle\}$ and
$\{q^{\prime}_i,|\phi_i\rangle\}$, respectively. The OD of the
tensor state $\rho_1\otimes\rho_2$ is then constituted by the
ensembles $\{p^{\prime}_iq^{\prime}_j,|\psi_i\phi_j\rangle\}$
because of $E_f(\rho_1\otimes\rho_2)=E_f(\rho_1)+E_f(\rho_2)$.
This makes the set
$\varepsilon_{\sigma}=\varepsilon_{\rho_1\otimes\rho_2}=\{r_{ij},|\psi_i\phi_j\rangle\}$
an OD family, too. When $E_f(\rho_1)$ and $E_f(\rho_2)$ are
additive, equivalently $E_f(\rho_1\otimes\rho_2)$ is additive by
(2) in lemma 1, the fact that the EOF of any state
$\sigma^{\prime}\in\varepsilon_{\sigma}$ is additive is
immediately derived from (1) in lemma 1.
\hspace*{\fill}$\blacksquare$

Proposition 1 is very useful for entanglement measures. In the
following text the known EOF of the states $\rho_i$'s on the
spaces $H_{A_i}\otimes H_{B_i}$'s respectively, helps generate the
OD families $\varepsilon_{\rho_i}=\{p_{ij},|\phi_{ij}\rangle\}$'s
such that the EOF of any state $\sigma_i\in\varepsilon_{\rho_i}$
is computable; that is,
$E_f(\sigma_i)=\sum_jp_{ij}E(|\phi_{ij}\rangle)$. Our method
indeed generates a family of states with computable EOF by using
only a state with known EOF. When $E_f(\sigma_i)$ is additive for
some state $\sigma_i$, by proposition 1 we can generate the new OD
families
$\varepsilon_{\sigma_i\otimes\sigma_j}=\{p_{ijk},|\phi_{ik}\rangle|\phi_{jk}\rangle\}$
on the space $H_{A_iA_j}\otimes H_{B_iB_j}$ and the EOF of any
state $\sigma_{ij}\in\varepsilon_{\sigma_i\otimes\sigma_j}$
similarly reads
$E_f(\sigma_{ij})=\sum_kp_{ijk}E(|\phi_{ik}\rangle|\phi_{jk}\rangle)$.
In the same vein, we can generate more OD families
$\varepsilon_{\sigma_i\otimes\sigma_j\otimes\cdots\otimes\sigma_k}$
on the space $H_{A_iA_j...A_k}\otimes H_{B_iB_j...B_k}$ and
calculate the EOF of states in it. When
$E_f(\sigma_i),E_f(\sigma_j),...,E_f(\sigma_k)$ are additive, by
proposition 1 they will lead to new states whose EOF is additive
and thus the computable entanglement cost.

For example, we consider the two-qubit MC state
\begin{equation}
\rho_{p,\theta}=p|\psi_{\theta}\rangle\langle\psi_{\theta}|+
(1-p)|\psi_{\pi/2-\theta}\rangle\langle\psi_{\pi/2-\theta}|,
\end{equation}
where
$|\psi_{\theta}\rangle=\cos\theta|00\rangle+\sin\theta|11\rangle$
and $|0\rangle,|1\rangle,...$ are computational basis. It follows
from \cite{Wootters,Vidal2} that
$E_f(\rho_{p,\theta})=E_c(\rho_{p,\theta})=h({\cos}^2\theta)$ with
the function $h(x)$ being the binary entropy function. So the set
$\varepsilon_{\rho_{p,\theta}}=\{p,1-p,|\psi_{\theta}\rangle,|\psi_{\pi/2-\theta}\rangle\}$
is an OD family. Thus by proposition 1 we can compute the EOF and
entanglement cost for any state $\rho$ in the OD family
$\varepsilon_{\rho_{p_1,\theta_1}\otimes\rho_{p_2,\theta_2}\otimes\cdots\otimes\rho_{p_n,\theta_n}}
=\{p_1,p_2,...,p_{2^n},
|\psi_{\theta_1}\psi_{\theta_2}...\psi_{\theta_{n}}\rangle,
|\psi_{\pi/2-\theta_1}\psi_{\theta_2}...\psi_{\theta_{n}}\rangle,...,\\
|\psi_{\pi/2-\theta_1}\psi_{\pi/2-\theta_2}...\psi_{\pi/2-\theta_{n}}\rangle\}$,
namely $E_f(\rho)=E_c(\rho)=\sum^{n}_{i=1}h({\cos}^2\theta_i).$
Remarkably, the EOF of different states in the OD family do not
change with the probability distributions. So we have proposed a
new family of MC states and thus $E_f(\rho)$ is additive. As the
distillable entanglement of the MC state is computable
\cite{Rains}, we thus obtain that the difference between the
entanglement cost and distillable entanglement of
$\rho\in\varepsilon_{\rho_{p_1,\theta_1}\otimes\rho_{p_2,\theta_2}\otimes\cdots\otimes\rho_{p_n,\theta_n}}$
on the space $H_{A}\otimes H_{B}$ is
\begin{equation}
E_c(\rho)-E_d(\rho)=\sum^{n}_{i=1}h({\cos}^2\theta_i)-S(\mbox{Tr}_A\rho)+S(\rho).
\end{equation}
This expression indeed represents the entanglement that cannot be
distilled from the state $\rho$ under LOCC \cite{Bennett1}. When
the conjecture that Eqs. (4) is strictly larger than zero is true,
it will mean the asymptotic entanglement manipulation of $\rho$ is
irreversible, which is essentially different from the case of pure
states \cite{Bennett2}. Much numerical calculation has been done
and they supported the conjecture. Furthermore, it turns out to be
true by plotting Eqs. (4), when the number of parameters in $\rho$
is less.

Similarly one can verify a more general two-qubit OD family
$\varepsilon_{\sigma}=\{q,1-q,\sqrt
p\cos\theta|00\rangle+\sqrt{1-p}\sin\theta
(x|00\rangle+y|01\rangle+z|11\rangle), \sqrt
p\sin\theta|00\rangle-\sqrt{1-p}\cos\theta
(x|00\rangle+y|01\rangle+z|11\rangle)\}$ up to the normalization
factors, $p,x,y,z\in[0,1], x^2+y^2+z^2=1$ and $
\tan\theta=\frac{-1+2(1-p)x^2-\sqrt{1+4(1-p)^2(x^4-x^2)}}{2x\sqrt{p-p^2}}.
$ The EOF of any state in $\varepsilon_{\sigma}$ is equal to
$h(\frac12+\frac12\sqrt{1-4(1-p)^2x^2z^2})$ and it is additive
\cite{Vidal2}. As the state $\rho_{p,\theta}$ in Eqs. (3) belongs
to $\varepsilon_\sigma$, the newly constituted OD family
$\varepsilon_{\sigma_1\otimes\sigma_2\otimes\cdots\otimes\sigma_n}$
provides more states $\rho$ whose EOF is computable and additive,
namely
$E_f(\rho)=E_c(\rho)=\sum^{n}_{i=1}h(\frac12+\frac12\sqrt{1-4(1-p_i)^2x^2_iz^2_i})$.
Generally, \cite{Wootters} has given a programmable way to find
out the OD of any two-qubit state $\rho$. It provides abundant
resource for the generation of new OD families, in which the EOF
of the states is computable by proposition 1.

In this case, we can obtain more states with computable EOF by
mixing $\varepsilon_{\rho}$ with other OD families, e.g., the
family $\varepsilon_s$ consisting of all separable states
$\rho_s$. As the entanglement cannot be increased under LOCC and
thus $E_f(\rho\otimes\rho_s)\geq E_f(\rho)$, it then follows from
the subadditivity of EOF that

\textit{Lemma 2.} The EOF of any state $\rho\in\varepsilon_s$ is
additive. The set of separable states $\varepsilon_s$ is an OD
family. \hspace*{\fill}$\blacksquare$

Although the separable state is classically established
\cite{Werner}, we emphasize that lemma 2 is by no means a trivial
result. We consider the $4\times4$ state $\sigma$ in the OD family
$\{p,1-p,|00\rangle|\psi_{\theta}\rangle,\frac{|0\rangle+|1\rangle}{\sqrt2}
\frac{|0\rangle+|1\rangle}{\sqrt2}|\psi_{\pi/2-\theta}\rangle\}$
\cite{notation2}, which is a subset of the OD family
$\varepsilon_{\rho_s\otimes\rho_{p,\theta}}$. So we get that
$E_f(\sigma)=E_c(\sigma)=h({\cos}^2\theta)$ and it is additive.
However, these results cannot be derived from any existing theory
since the partial trace over a system of $\sigma$ does not equal
to an entanglement-breaking channel \cite{Vidal2}. In addition,
the tag states $|00\rangle$ and
$\frac{|0\rangle+|1\rangle}{\sqrt2}
\frac{|0\rangle+|1\rangle}{\sqrt2}$ cannot be explicitly
distinguished \cite{Horodecki2}. Generally given an OD family
$\varepsilon_{\rho}=\{p_i,|\psi_i\rangle\}$, by proposition 1 one
can always constitute the new OD family
$\varepsilon_{\rho_s\otimes\rho}=\{p_i,|\phi_i\rangle|\psi_i\rangle\}$
with each state $|\phi_i\rangle$ in randomly product form. The EOF
of any state $\sigma\in\varepsilon_{\rho_s\otimes\rho}$ thus reads
$E_f(\sigma)=\sum_ip_iE(|\psi_i\rangle)$. This value will coincide
with $E_c(\sigma)$ when $E_f(\rho)$ is additive.

We have known that the OD of the two-qubit state $\rho$ is
constituted by pure states with identical amount of entanglement
\cite{Wootters}. However it is not the case for states of high
dimensions, as the following example illustrates. Suppose
$|\psi\rangle=\sum^{d-1}_{i=0}c_i|ii\rangle,
|\phi\rangle=(\sum^{f-1}_{i=0}c^2_i)^{-1/2}\sum^{f-1}_{i=0}c_i|ii\rangle,
\cos\theta=(\sum^{f-1}_{i=0}c^2_i)^{1/2},
\sin\theta=(\sum^{d-1}_{i=f}c^2_i)^{1/2},
u_{00}=-u_{11}=\sqrt{\frac12+\frac{-1+p+p\cos2\theta}{2\sqrt{1-p^2\sin^22\theta}}},
u_{01}=u_{10}=\sqrt{\frac12-\frac{-1+p+p\cos2\theta}{2\sqrt{1-p^2\sin^22\theta}}},
\forall c_i>0, d>f,$ we have

\textit{Lemma 3.} Consider the MC state
$\rho=p|\psi\rangle\langle\psi|+(1-p)|\phi\rangle\langle\phi|$.
The OD and EOF of $\rho$ respectively read
$
\sum^{1}_{i=0}(u_{i0}\sqrt{p}|\psi\rangle+u_{i1}\sqrt{1-p}|\phi\rangle)
(u_{i0}\sqrt{p}\langle\psi|+u_{i1}\sqrt{1-p}\langle\phi|)
$
and
\begin{eqnarray}
&E_f(\rho)=-p\sum\limits^{d-1}_{i=0}c^2_i\log{c^2_i}
-(1-p)\sum\limits^{f-1}_{i=0}\frac{c^2_i}{\cos^2\theta}\log{\frac{c^2_i}{\cos^2\theta}}\nonumber\\
&+h\big(\frac12+\frac12\sqrt{1-p^2\sin^22\theta}\big)-p\
h(\cos^2\theta).
\end{eqnarray}
\textit{Proof.} It follows from the definition of EOF
\cite{Wootters} that
\begin{equation}
E_f(\rho)=\underset{\vec{v}_0,\vec{v}_1}{\mbox{min}}\sum\limits_i
q_iE\bigg(\frac{v_{i0}\sqrt{p}|\psi\rangle+v_{i1}\sqrt{1-p}|\phi\rangle}{\sqrt{q_i}}\bigg),
\end{equation}
where $[v_{ij}]=[\vec{v}_0,\vec{v}_1]$ are the first two columns
of a unitary matrix and $\sqrt{q_i}$ is the normalization factor.
Notice every decomposition of $\rho$ is expressed in Eqs. (6)
owing to the linear independence of $|\psi\rangle$ and
$|\phi\rangle$. Reduction of Eqs. (6) leads to the optimization
problem equivalent to that for the two-qubit MC state, whose
entanglement has been derived by \cite{Wootters}. One can check
that the entanglement of the proposed OD of $\rho$ coincides with
$E_f(\rho)$. \hspace*{\fill}$\blacksquare$

Lemma 3 actually provides a new family of MC states with
computable EOF. One can easily check that the amounts of
entanglement contained in the two pure states constituting
$\varepsilon_\rho$ are always not equal, which totally differs
with the OD families of two-qubit states and separable states. It
thus by proposition 1 generates new OD family
$\varepsilon_{\rho_1\otimes\rho_2\otimes\cdots\otimes\rho_n}$ in
which the pure states have different entanglement, correspondingly
the states in
$\varepsilon_{\rho_1\otimes\rho_2\otimes\cdots\otimes\rho_n}$ are
not equally entangled when the probability distributions change.
So this OD family generates more complex states with computable
entanglement cost. Of course more OD families can be produced by
using the results on two-qubit states and separable states.

Next, we show that there is indeed a gap between the distillable
entanglement and entanglement cost for $\rho$. It follows from
\cite{Rains} and Eqs. (5) that
\begin{eqnarray}
&E_c(\rho)-E_d(\rho)=
h\big(\frac12+\frac12\sqrt{1-p^2\sin^22\theta}\big)-h(p\sin^2\theta)\nonumber\\
&+h\big(\frac12+\frac12\sqrt{1-4p\sin^2\theta+4p^2\sin^2\theta}\big),
\end{eqnarray}
which turns out to be strictly larger than zero unless $p=0$ or 1.
A simple approach to this inequality is by plotting Eqs. (7).
Differing with the qualitative derivation (e.g.,
\cite{Vollbrecht1}), our result analytically shows that one can
distill the same entanglement from $\rho$ as $E_c(\rho)$ if and
only if $\rho$ is pure. The undistillable entanglement is
explicitly given by Eqs. (7). It strongly supports the conjecture
that entanglement distillation is irreversible in general
\cite{Horodecki3}.

Another proof for OD family containing different entanglement
concerns an important state in QIT, the isotropic state
$\rho_F=\frac{1-F}{d^2-1}I+\frac{Fd^2-1}{d^2-1}|\psi^+\rangle\langle\psi^+|,
|\psi^+\rangle=\frac{1}{\sqrt d}\sum^{d}_{i=1}|ii\rangle$
\cite{Horodecki4}. It turned out that for $\rho_F$ with $d=3$ and
$F>8/9$, $\varepsilon_{\rho_F}$ consists of a maximally entangled
state and nine states $\rho$ obtained by twirling and they are
thus equivalently entangled, i.e., $E(\rho)=-1/3+\mbox{log} \ 3$
\cite{Terhal2}. We prove that $\varepsilon_{\rho_F}$ of high
dimension $d,F>(4d-4)/d^2$ is similarly composed of
$|\psi^+\rangle$ and many states $\rho$ obtained by twirling, each
of which contains entanglement
$E(\rho)=\frac{2-d}d\mbox{log}(d-1)+\mbox{log}\ d$. This
conclusion follows from the OD of $\rho_F$ such that
\begin{eqnarray}
\rho_F&=&\frac{d^2(1-F)}{(d-2)^2}\sum^{L}_{l=1}\frac1L|\psi_l\rangle\langle\psi_l|
+\frac{4-4d+Fd^2}{(d-2)^2}|\psi^+\rangle\langle\psi^+|,\nonumber\\
|\psi_l\rangle&=&\frac{d-2}{\sqrt{d^2-d}}\sum^{d}_{i=1}a_{li}|i\rangle\sum^{d}_{i=1}a^*_{li}|i\rangle
+\frac{1}{\sqrt{d^2-d}}\sum^{d}_{i=1}|ii\rangle.
\end{eqnarray}
We demonstrate the $L\times d$ coefficient matrix $[a_{li}]$ in
the case of odd $d\geq3$. We regard the $n\times1$ vectors
$\vec{a}_j,j=1,...,d$ as the nonzero entries in $[a_{li}]$, each
``row" of which consists of $d+1/2$ nonzero entries and $d-1/2$
zero. The subscript of $\vec{a}_j$ marks its column in $[a_{li}]$.
The first row of $[a_{li}]$ consists of
$[\vec{a}_1,...,\vec{a}_{d+1/2},0,...,0]$, the second
$[\vec{a}_1,...,\vec{a}_{d-1/2},0,\vec{a}_{d+3/2},0,...,0],...,$
and the last $[0,...0,\vec{a}_{d+1/2},...,\vec{a}_{d}]$. Counting
all kinds of combinations, there are in all ${d\choose d+1/2}$
rows and hence $L=n{d\choose d+1/2}$. Each $\vec{a}_j$ has the
form
$(\sqrt{\frac{2}{d+1}},\sqrt{\frac{2}{d+1}}e^{f_j\frac{2\pi}{n}i},...,\\
\sqrt{\frac{2}{d+1}}e^{f_j\frac{2\pi}{n}(n-1)i})^T$, where the
natural numbers $f_j$'s are required to be $n|f_i+f_j-f_k-f_l$ if
and only if $i=k,j=l$ or $i=l,j=k.$ It can be done, e.g., by
choosing $f_i=m^i,m>1,n>2m^d-4$. One can deal with the case of
even $n$ similarly and verify Eqs. (8). So we have given the
method of generating numerous $\varepsilon_{\rho_F}$, which is a
remarkable character of the isotropic state. It makes
$\varepsilon_{\rho_F}$ a much stronger OD family compared to the
former results, when it is used to create the states with
computable EOF by proposition 1. Specially, the EOF of any state
in $\varepsilon_{\rho_F}$ has the form
$p(\frac{2-d}d\mbox{log}(d-1)+\mbox{log}\ d)+(1-p)\mbox{log}\ d$
with the probability $p\in[0,1]$. In addition, the OD of $\rho_F$
implies that one can generate the isotropic states by classically
mixing the ensemble of MC states
$\rho=p|\psi_l\rangle\langle\psi_l|+(1-p)|\psi^+\rangle\langle\psi^+|$
obtained by twirling.

On the other hand there exist the states of high dimensions whose
OD consists of equivalently entangled states, e.g., the Werner
state $\rho_w$, which is important in QIT and has been extensively
investigated \cite{Werner,Lee}. A $d\times d$ Werner state has the
form $ \rho_w=\frac{d-F}{d^3-d}I+\frac{d\
F-1}{d^3-d}\sum^{d-1}_{i,j=0}|ij\rangle\langle ji|$. The EOF of an
entangled Werner state ($F\in[-1,0)$) has the analytical
expression $E_f(\rho_w)=h(1/2+\sqrt{1-F^2}/2)$ \cite{Vollbrecht2}.
From it we can derive the OD of $\rho_w$ such that
\begin{eqnarray}
&\rho_w=\frac{1}{2d^2-2d}\sum\limits^{d-1}_{i>j=0}\sum\limits^3_{k=0}
|\psi_{ijk}\rangle\langle\psi_{ijk}|,\nonumber\\
&|\psi_{ijk}\rangle=2u_{k0}\sqrt{\frac{F+1}{2d+2}}|ii\rangle
+2u_{k3}\sqrt{\frac{1-F}{2}}\frac{|ij\rangle-|ji\rangle}{\sqrt2}\nonumber\\
&+2u_{k1}\sqrt{\frac{F+1}{2d+2}}|jj\rangle
+2u_{k2}\sqrt{\frac{(d-1)(F+1)}{2d+2}}\frac{|ij\rangle+|ji\rangle}{\sqrt2},\nonumber\\
&[u_{ij}]=\left(\begin{array}{cccc}
-1/2 & 1/2 & 1/2 & 1/2\\
1/2 & -1/2 & 1/2 & 1/2\\
i/2 & i/2 & -1/2 & 1/2\\
i/2 & i/2 & 1/2 & -1/2
\end{array}\right).
\end{eqnarray}
One can easily verify the proposed OD which indeed gives rise to
the Werner-OD family
$\varepsilon_w=\{p_{ijk},|\psi_{ijk}\rangle\}$, and each pure
state $|\psi_{ijk}\rangle$ has identical amount of entanglement
$E_f(\rho_w)$. It thus implies that the EOF of any state in
$\varepsilon_w$ is equal to $E_f(\rho_w)$, no matter how the
probability distributions change. This is similar to the case of
two-qubit states. We can use the Werner-OD family to construct
more OD families with computable EOF by proposition 1. It is a new
function of the Werner state. Moreover, one can check that any
state $\sum_{ij}p_{ij}\rho_{ij}\in\varepsilon_w$ is negative
partial transpose (NPT). This helps infer the EOF and
irreversibility of NPT bound entangled states, if it really exists
\cite{DiVincenzo}.

We notice that for each pair of fixed $i,j$, the state $\rho_{ij}
\in\varepsilon_w$ is just a $2\times2$ Werner state $\rho_{w0}$.
Such an OD is interesting in the sense that we can experimentally
prepare a $d\times d$ Werner state by means of classically mixing
many states $\rho_{w0}$'s up to unitary operations with identical
probabilities $p_{ij}$'s, equivalently making a state $\rho_{w0}$
go through the unital channel
$\Lambda(\rho)=\frac{2}{d^2-d}\sum^{d-1}_{i>j=0}v_{ij}\otimes
w_{ij}\rho_{w0}v^{\dag}_{ij}\otimes w^{\dag}_{ij}$. Here each pair
of unitary operations $v_{ij}$ and $w_{ij}$ acts on a $2\times2$
space, so they can be indeed regarded as Pauli operations $I$ and
$\sigma_x$. As the state $\rho_{w0}$ has been realizable
\cite{Lee}, the proposed experiment is probably realizable by
current techniques. More importantly, the experiment will verify
that the $E_f(\rho_w)$ is indeed the minimal entanglement required
to create a Werner state of high dimension.

To summarize, we have presented the OD method of deriving the EOF,
additivity and entanglement cost for many states. Our method is
flexible and could yield more results on entanglement measures. It
also helps generate the Werner state and check the EOF
experimentally.

We thank D. Yang for very useful discussions. The work was partly
supported by the NNSF of China Grant No.90503009, No.10775116, and
973 Program Grant No.2005CB724508.


\begin{references}
\bibitem{notation2} We choose the form so that checking the
entanglement-breaking channel \cite{Vidal2} is easy.
\end{references}
\end{document}